\documentstyle[emulateapj,psfig]{article}
\begin{document}

\slugcomment{Accepted for Publication in the Astrophysical Journal Letters}
 
\title{Hydrodynamics of Relativistic Blast Waves in a Density-Jump Medium
and Their Emission Signature}
\author{Z. G. Dai and T. Lu}
\affil{Department of Astronomy, Nanjing University, Nanjing 210093, China\\
E-mail: daizigao@public1.ptt.js.cn; tlu@nju.edu.cn}

\begin{abstract}
We analyze in detail the hydrodynamics and afterglow emission of
an ultrarelativistic blast wave when it expands in a density-jump medium.
Such a medium is likely to appear in the vicinity of gamma-ray bursts (GRBs)
associated with massive stars. The interaction of the blast wave with
this medium is described through a reverse shock and a forward shock.
We show that the reverse shock is initially relativistic if the factor of a density
jump ($\alpha$) is much larger than 21, and Newtonian if $1<\alpha\ll 21$. 
We also calculate light curves of the afterglow emission during the interaction 
if  the reverse shock is relativistic, and find that the optical flux density initially 
decays abruptly, then rises rapidly, and finally fades based on a power-law, 
which could be followed by an abrupt decay when the reverse shock has just 
crossed the originally swept-up matter. Therefore, one property of an afterglow
occurring in a large-density-jump medium is an abrupt drop followed
by a bump in the light curve and thus provides a probe of circumburst 
environments. In addition, this property could not only account for the optical 
afterglows of GRB 970508 and GRB 000301C but also explain the X-ray afterglow
of GRB 981226.  
\end{abstract}


%%%%%%%%%%%%%%%%%%%%%%%%%%%%%%%%%%%%%%%
\section {Introduction}

The density distribution of circumburst environments is one of the most 
important issues in the theories of gamma-ray bursts (GRBs) (for review 
articles see Piran 1999; van Paradijs, Kouveliotou \& Wijers 2000; 
M\'esz\'aros 2001). On one hand, it is directly pertinent to the progenitors
of GRBs. Two currently popular models for the progenitors are the mergers of
compact stars (neutron stars or black holes) and the explosions of massive stars.
It has been argued that GRBs produced by the former model occur in a
uniform interstellar medium (ISM) with density of $\sim 1\,{\rm cm}^{-3}$ and
GRBs in the latter model occur in pre-burst winds (Chevalier \& Li 1999) and/or
giant molecular clouds (Galama \& Wijers 2001; Reichart \& Price 2001). Thus,
an environmental signature may provide a clue about the GRB progenitors.
On the other hand,  the environmental properties can directly
influence the decay rates of afterglows. For example, afterglows arising
from the interaction with pre-burst winds should decay more rapidly
than afterglows do in a low-density medium (e.g., ISM) (Dai \& Lu 1998a;
M\'esz\'aros, Rees \& Wijers 1998; Panaitescu, M\'esz\'aros \& Rees 1998;
Chevalier \& Li 1999, 2000). Furthermore, ultrarelativistic fireballs (or jets)
in a uniform dense medium (e.g, galactic-like giant molecular clouds) must
evolve to the non-relativistic regime within a few days after the bursts, leading
to a rapid decay of the afterglows (Dai \& Lu 1999, 2000; Wang, Dai \& Lu 2000).
It is thus natural that an afterglow signature can probe the ambient matter
as well as the progenitors.

In the previous afterglow shock models, the environments of GRBs are usually
assumed to be continuous media (e.g., ISM and wind). Actually, there are
possibly jumps (or bumps) in the density profile of the ambient media of GRBs
associated with massive stars. Such jumps may be produced by several
astrophysical processes, e.g., the deceleration of winds in their external medium 
(Ramirez-Ruiz et al. 2001; R. Wijers 2001, private communication) or the interaction
of fast and slow winds (Luo \& McCray 1991; Vikram \& Balick 1998) or
pre-burst supernova ejecta (Vietri \& Stella 1998). In this Letter, we perform
a careful analysis for the afterglow emission when a post-burst relativistic
blast wave interacts with such a density-jump medium.  We first analyze
the hydrodynamics of the interaction in \S 2 and then discuss the afterglow
signature in detail in \S 3. Our findings are summarized and discussed in \S 4.

%%%%%%%%%%%%%%%%%%%%%%%%%%%%%%%%%%%%%%%%%% 
\section{Hydrodynamics}

Let's envision that some central energy source produces an ultrarelativistic
fireball. After the internal shock emission (viz., a GRB), the fireball will start to
sweep up its ambient medium, leading to an ultrarelativistic blast wave. The
initial hydrodynamics of this interaction has been analyzed in detail by Sari \&
Piran (1995), and studied numerically by Kobayashi, Piran \& Sari (1999) and
Kobayashi \& Sari (2000). Now we consider this ultrarelativistic blast wave 
which first expands in an ISM or a stellar wind and then hits an outer high-density 
region. We assume that the medium of interest has a simple density profile: 
$n=AR^{-s}$ for $R\le R_0$ and $n=n_1={\rm constant}$ for $R>R_0$. 
Here $A=n_0\times 1\,{\rm cm}^{-3}$ if the inner medium is an ISM ($s=0$), 
and $A=3\times 10^{35}A_*\,{\rm cm}^{-1}$ if the inner medium is a wind ($s=2$). 
Such  a density profile seems to be able to reconcile the 
contradiction that the multiwavelength afterglow fits in the jet model 
by Panaitescu \& Kumar (2001) indicate an ambient medium density significantly 
lower than that expected in star-forming regions. Ramirez-Ruiz et al. (2001) 
showed that the interaction of winds from massive stars with the ambient medium 
can indeed lead to a large density jump at a few $10^{17}$ cm.  

Before the blast wave hits the high-density medium, its Lorentz factor decays
based on the Blandford-McKee's (1976) solution as $\gamma=8.2E_{53}^{1/8}
n_0^{-1/8}t^{-3/8}$ in the inner-ISM case and $\gamma=8.8E_{53}^{1/4}
A_*^{-1/4}t^{-1/4}$ in the inner-wind case, where $E_{53}$ is the energy of
the blast wave in units of $10^{53}$ ergs and $t$ is the observer's time in units
of 1 day (neglecting the redshift effect). Subsequently,
the interaction of the blast wave with the high-density medium is
described through two shocks: a reverse shock that propagates into
the hot shell (viz., the $R\le R_0$ medium swept up by the blast
wave), and a forward shock that propagates into the high-density
medium. Therefore, there are four regions separated
by the two shocks in this system: (1) unshocked high-density
medium, (2) forward-shocked high-density medium, (3) reverse-shocked hot shell,
and (4) unshocked hot shell. We denote $n_i$, $e_i$ and $p_i$ as the baryon
number density, energy density and pressure of region ``$i$" in its own rest
frame respectively; $\gamma_i$ and $\beta_i$ are the Lorentz factor and
dimensionless velocity of region ``$i$" measured in the local medium's rest frame
respectively; and $\gamma_{ij}$ and $\beta_{ij}$ are the relative Lorentz factor and
dimensionless velocity of region ``$i$" measured in the rest frame of region
``$j$" respectively. If $\gamma_i\gg 1$ and $\gamma_j\gg 1$, then
$\gamma_{ij}\simeq (\gamma_i/\gamma_j+\gamma_j/\gamma_i)/2$.
We further assume the equations of state for regions 2, 3 and 4  to be
relativistic and region 1 to be cold. Thus, the equations describing the jump
conditions for the forward and reverse shocks become (Blandford \& McKee 1976;
Sari \& Piran 1995; Kumar \& Piran 2000; Zhang \& M\'esz\'aros 2001a)
\begin{equation}
\frac{e_2}{n_2m_pc^2}=\gamma_2-1,  \,\,\,\,\,\,
\frac{n_2}{n_1}=4\gamma_2+3,
\end{equation}
\begin{equation}
\gamma_{34}^2=\frac{(1+3e_3/e_4)(3+e_3/e_4)}{16e_3/e_4},
\end{equation}
\begin{equation}
\left(\frac{n_3}{n_4}\right)^2=\frac{(e_3/e_4)(1+3e_3/e_4)}{3+e_3/e_4},
\end{equation}
where $m_p$ is the proton mass.

Regions 2 and 3 should keep the pressure equilibrium and velocity equality
along the contact discontinuity, which yield $\gamma_2=\gamma_3$ and
$e_2=e_3$. Under these conditions, the solution of equations (1)-(3) depends
only on two parameters: $\gamma_4$ and $f\equiv e_4/(n_1m_pc^2)$.
The solution has two limits which correspond to the cases that
the reverse shock is relativistic or Newtonian. If $e_3\gg e_4$,
then the reverse shock is initially relativistic:
\begin{equation}
\gamma_2=\gamma_3=\frac{\gamma_4^{1/2}f^{1/4}}{3^{1/4}}, \,\,\,\,\,\,
\gamma_{34}=\frac{3^{1/4}\gamma_4^{1/2}}{2f^{1/4}}\gg 1,
\end{equation}
which requires $\alpha\equiv n_1/n_0\gg 64/3\simeq 21$,
where $n_0$ is the baryon number density of the inner medium at $R=R_0$ and 
the energy density of region 4 at this radius is assumed to equal 
$4\gamma_4^2n_0m_pc^2$. From equation (4), $\gamma_3\ll \gamma_4$, 
showing that most of the initial kinetic energy of region 4 is converted into 
thermal energy by the shocks. On the other hand, for $1<\alpha\ll 21$, 
the reverse shock is Newtonian:
\begin{equation}
\gamma_{34}-1\simeq \frac{1}{2}\left(\frac{2\gamma_4/\sqrt{f}-1}{2\gamma_4
/\sqrt{f}+2/\sqrt{3}}\right)^2\equiv \frac{1}{2}\xi^2, 
\end{equation}
\begin{equation}
\gamma_2=\gamma_3\simeq \gamma_4(1-|\xi|).
\end{equation}
In this case, the reverse shock converts only a small fraction ($|\xi|\ll 1$) of the
kinetic energy into thermal energy because $2\gamma_4\sim \sqrt{f}$. Thus,
the forward shock expands almost at the velocity of the previous blast wave.

In the next section, we will discuss light curves of the afterglow
emission if the reverse shock is initially relativistic. For this purpose, we need to
know how the thermodynamic quantities and Lorentz factor of each region
evolve with radius $R$ at two different stages:

When the reverse shock crosses region 4, this region always expands
adiabatically at a constant Lorentz factor of $\gamma_4$, and thus
we have $n_4\propto R^{-3}$ and $e_4\propto n_4^{4/3}\propto R^{-4}$.
As a result, we obtain $f\propto R^{-4}$. For regions 2 and 3, we have
$\gamma_2=\gamma_3\propto f^{1/4}\propto R^{-1}$, $e_2=e_3\propto
\gamma_2^2\propto R^{-2}$, $n_2\propto\gamma_2\propto R^{-1}$, and
$n_3\propto n_4(e_3/e_4)^{1/2}\propto R^{-2}$. At this stage, 
$\gamma_{34}\propto R$, which is different from the initial hydrodynamics of 
a relativistic blast wave, $\gamma_{34}\propto R^{3/4}$, derived from 
Sari \& Piran (1995) in the thin shell case. This is due to a hot region 4
as compared to Sari \& Piran (1995).

After the reverse shock crosses region 4, the profile of the shocked high-density
medium begins to approach the Blandford-McKee solution as long as the shocked
matter has a relativistic equation of state, as shown numerically
at the initially hydrodynamic stage of an afterglow by Kobayashi et al. (1999) and
Kobayashi \& Sari (2000). Using this solution, the Lorentz factor and the energy
density of a given fluid element for region 3 decay as $\gamma_3\propto R^{-7/2}$
and $e_3\propto R^{-26/3}$ (Sari \& Piran 1999a, 1999b). In addition, the Lorentz
factor and the energy density of region 2 evolve as $\gamma_2\propto R^{-3/2}$
and $e_2\propto R^{-3}$.

%%%%%%%%%%%%%%%%%%%%%%%%%%%%%%%%%%%%%%%%%
\section{Light Curves of the Emission}

We now consider synchrotron radiation from all the regions at two different
stages. The electron energy distribution just behind the shock is usually
a power-law: $dn_e/d\gamma_e\propto \gamma_e^{-p}$ for $\gamma_e\ge
\gamma_m$. Here we discuss only the case of $p>2$. Dai \& Cheng (2001)
have discussed light curves of the emission from a relativistic shock
in an ISM or a wind for $1<p<2$. Assuming that $\epsilon_e$ and $\epsilon_B$
are constant fractions of the internal energy density going into the electrons
and the magnetic field respectively, we have the electron minimum Lorentz
factor, $\gamma_{m,i}=[(p-2)/(p-1)]\epsilon_e e_i/(n_im_ec^2)$, and
the magnetic field, $B_i=(8\pi\epsilon_B e_i)^{1/2}$, for region ``$i$", 
where $m_e$ is the electron mass.

According to Sari, Piran \& Narayan (1998), the spectrum consists of four
power-law parts with three break frequencies: the self-absorption frequency,
the typical synchrotron frequency $\nu_{m,i}= \gamma_i\gamma_{m,i}^2
eB_i/(2\pi m_ec)$, and the cooling frequency $\nu_{c,i}=18\pi em_ec/(\sigma_T^2
B_i^3\gamma_it^2)$, where $\sigma_T$ is the Thomson scattering cross section.
In this Letter we neglect the self-absorption because it does  not affect
the optical radiation which we are interested in. In order to calculate
the flux density at a fixed frequency, one still needs to derive
the peak flux density. The observed peak flux density is given by
$F_{\nu_m,i}= N_{e,i}\gamma_iP_{\nu_m,i}/(4\pi D_L^2)$, where $N_{e,i}$ is
the electron number of region ``$i$" at radius $R$,
$P_{\nu_m,i}=m_ec^2\sigma_TB_i/(3e)$ is the radiated power per electron
per unit frequency in the frame comoving with the shocked matter,  and
$D_L$ is the source's luminosity distance to the observer. So we in fact need
to calculate $N_{e,i}$. First, it is easy to obtain the total electron number of 
region 2 by $N_{e,2}=(4\pi/3)n_1(R^3-R_0^3)$. Second, the time interval for 
the reverse shock to spend in crossing a length interval $dx'_4$ in the rest frame 
of region 4 is $dt'_4=dx'_4/(\beta_{34}c)$, which is related to a time interval 
($dt'_3$) in the rest frame of region 3 by $dt'_4=\gamma_{34}dt'_3$, and 
furthermore $dt'_3$ is related to a time interval ($dt_{\rm m}$) in the local medium's 
rest frame by $dt'_3=dt_{\rm m}/\gamma_3=dR/(\gamma_3\beta_3c)$, 
so we calculate the electron number of region 3:
$N_{e,3}=\int_0^{x'_4}4\pi R^2n_4dx'_4=\int_{R_0}^R4\pi R^2n_4
[\gamma_{34}\beta_{34}/(\gamma_3\beta_3)]dR
\propto (R^2-R_0^2)$, where $\beta_{34}\approx \beta_3\approx 1$, 
and thus the electron number of region 4 is $N_{e,4}=4\pi AR_0^{3-s}/(3-s)-
N_{e,3}$, when the reverse shock crosses region 4. Letting $N_{e,3}$
equal the initial electron number of region 4 (viz., when $N_{e,4}=0$), we can 
calculate the radius at which the reverse shock has just crossed this region:
$R_\Delta  =  R_0(1+5.83\times 10^{-3}n_0^{1/2}n_{1,3}^{-1/2})$ for $s=0$
and $R_\Delta  =  R_0(1+1.25\times 10^{-2}E_{53}^{-1/2}A_*n_{1,3}^{-1/2}
t_0^{-1/2})$ for $s=2$,
where $n_{1,3}=n_1/10^3{\rm cm}^{-3}$ and $t_0$ is the observer's time (in 1 day)
at $R=R_0$.

Although the spectrum does not depend on the hydrodynamics of the shocked
matter, the light curve at a fixed frequency is determined by the temporal
evolution of $\nu_{m,i}$, $\nu_{c,i}$ and $F_{\nu_m,i}$. These quantities depend on
how $\gamma_i$, $n_i$, $e_i$ and $N_{e,i}$ scale as a function of $R$
as well as $t$ either for $R_0\le R\le R_\Delta$ or for $R>R_\Delta$. However,
we note that for the typical values of the involved parameters (see below),
$R_\Delta/R_0-1\ll 1$, showing that $\gamma_i$, $n_i$ and $e_i$ are almost
unchanged and thus we consider only the temporal evolution
of $\nu_{c,i}$ ($\propto t^{-2}$) and $N_{e,i}$ for $R_0\le R\le R_\Delta$.

One crucial effect in calculating the light curve is that the photons 
which are radiated from different regions at the same time measured
in the local medium's rest frame will be detected at different observer
times (Zhang \& M\'esz\'aros 2001a). 
The understanding of this effect is that for $R_0\le R\le R_\Delta$ 
the Lorentz factor of region 4 is much larger than that of regions 2 and 3 
so that for a same time interval in the local medium's rest frame, $dR/c$, 
the emission from region 4 reaches the observer in a time interval of 
$\sim dR/(2c\gamma_4^2)$, while the emission from regions 3 and 2
reaches the observer in a time interval of $\sim dR/(2c\gamma_3^2)$. 
After considering this effect and the scaling law of $\gamma_i$ with $R$, 
we obtain the observer time of the radiation from region 4 at $R=R_\Delta$:
$t_{\Delta,4}=t_0(1+1.16\times 10^{-2}n_0^{1/2}n_{1,3}^{-1/2})$ 
for $s=0$ and $t_{\Delta,4}=t_0(1+2.50\times 10^{-2}E_{53}^{-1/2}
A_*n_{1,3}^{-1/2}t_0^{-1/2})$ for $s=2$,
and the corresponding observer time of the radiation from regions 3 and 2:
$t_{\Delta,3}=1.33t_0$ for $s=0$ and $t_{\Delta,3}=2.0t_0$ for $s=2$,
where we have used $R_\Delta/R_0-1\ll 1$. It is thus seen 
that $t_{\Delta,4}/t_0-1\ll t_{\Delta,3}/t_0-1$, implying that the observed
radiation from regions 3 and 2 is indeed delayed as compared with that from
region 4.

Fig. 1 presents two R-band ($\nu_R\simeq 4.4\times 10^{14}$ Hz)
light curves about the effect of an ultrarelativistic blast wave
interacting with a density-jump medium on the afterglow.
The outer-medium density assumed here, $n_1\sim 10^3\,{\rm cm}^{-3}$, 
is the one of typical galactic-like giant molecular clouds. It can be seen from 
this figure that at $t\ge t_0$ the flux density ($F_{\nu_R}$) initially drops abruptly, 
then rises rapidly, and finally declines based on a power-law followed 
by an abrupt decay at $t=t_{\Delta,3}$.  This result is easily understood:
an initially abrupt decay of the emission is due to the spectral cutoff frequency 
$\nu_{{\rm cut},4}<\nu_R$ ({\em solid line}) or the rapid decrease of 
the electron number in region 4, $N_{e,4}\propto [4t_{\Delta,4}/t_0-(t/t_0+1)^2]$ 
({\em dashed line}), during the adiabatic expansion for $t_0\le t\le t_{\Delta,4}$. 
Note that $\nu_{{\rm cut},4}=\nu_{c,0}(R/R_0)^{-4}$ results from evolution of 
the fast-cooling electrons of region 4, where $\nu_{c,0}=2.8\times 10^{13}
\epsilon_{B,-1}^{-3/2}E_{53}^{1/2}n_0^{-1}t_0^{-1/2}\,{\rm Hz}$ is 
the cooling frequency of the originally swept-up electrons at $t=t_0$. 
In the period of $t_0\le t\le t_{\Delta,4}$, the emission flux densities from regions 
2 and 3 are low both because the two shocks have swept up only a small number 
of the electrons and because the radiation from these regions reaches
the observer at a later time than the radiation from region 4
does.  As the number of the electrons swept up by the two shocks
increases, the flux density increases rapidly as $F_{\nu_R}\propto
(t/t_0-1)(t/t_0)^{-1}$, where the first factor arises from $F_{\nu_m,i}\propto
(t/t_0-1)$ and the second factor from $\nu_{c,i}\propto (t/t_0)^{-2}$
for regions 2 and 3. However, since $\nu_{c,3}< \nu_{m,3}<\nu_R$
for the parameters shown in the figure, all the electrons in region 3 
are in the fast cooling regime. As a result, the R-band flux density 
of the radiation from this region disappears at $t\ge t_{\Delta,3}$ and 
thus only the radiation from region 2 could be detected. Therefore, 
an abrupt decay of the flux density could appear at $t=t_{\Delta,3}$.
We further define the factor of emission brightening ($\cal{R}$) as
the ratio of the observed density fluxes with and without a density jump
at $t=t_{\Delta,3}$. In Fig. 1, ${\cal R}\sim 5.1$ ({\em solid line}) and
${\cal R}\sim 10$ ({\em dashed line}). Fig. 2 exhibits the light curves of 
the afterglow emission when the inner medium at $R\le R_0$ is a stellar wind. 
In this figure, ${\cal R}\sim 5.3$ ({\em solid line}) and ${\cal R}\sim 11$ 
({\em dashed line}). Recently Ramirez-Ruiz et al. (2001) also estimated 
${\cal R}\sim \alpha^{(p+1)/[4(4-s)]}\simeq 4.5$ and $21$ for $s=0$ and 
$s=2$ (where $p=2.5$ and $\alpha=10^3$) respectively. However, we find 
that $\cal R$ depends on the shock parameters (e.g, $\epsilon_e$ and 
$\epsilon_B$) for a fixed density jump. One reason for the difference between 
our and Ramirez-Ruiz et al.'s (2001) results is that for $t_0<t<t_{\Delta,3}$ 
the optical emission arises partially from the reverse shock considered here. 
Another reason is that the electrons which produce the optical emission 
may be in different radiation regimes before and when the reverse shock 
crosses region 4. 
 
%%%%%%%%%%%%%%%%%%%%%%%%%%%%%%%%%%%%%%%%
\section{Discussion and Conclusions}

We have analyzed the hydrodynamics and afterglow emission of
an ultrarelativistic blast wave when it interacts with a density-jump medium.
This interaction is described through two shocks: a reverse shock and
a forward shock. We have shown that the reverse shock is initially relativistic
if $\alpha\gg 21$, and Newtonian if $1<\alpha \ll 21$. We have also investigated 
in detail light curves of the afterglow emission during the interaction 
if  the reverse shock is relativistic, and found that the R-band flux
density initially decays abruptly, then rises rapidly, and finally fades
based on a power-law, which could be followed by an abrupt decay
when the reverse shock has just crossed the originally swept-up matter.

Our analysis is based on several simplifications: First, we have considered
only one density jump, but our discussion should be in principle applied to
a more realistic case in which there could be a few jumps (or bumps) in the 
density profile. Second, since the $Y$ parameter for synchrotron self-Compton
(SSC) is not far larger than unity in our model, the effect of SSC
on the optical emission is insignificant. This effect was recently
discussed by several authors (e.g., Waxman 1997; Wei \& Lu 1998;
Panaitescu \& Kumar 2000; Sari \& Esin 2001; Zhang \& M\'esz\'aros 2001b).
Third, we have assumed a spherical relativistic blast wave, but an actual
blast wave may be a jet, whose edge effect and sideways expansion can
lead to a steepening of the light curve (M\'esz\'aros \& Rees 1999;
Rhoads 1999; Sari, Piran \& Halpern 1999; Dai \& Cheng 2001). Fourth, we 
have neglected the emission off the line of sight.  Such emission could 
significantly alleviate the initial abrupt drop of the flux density.   

The humps have been observed to appear in the light curves
of several optical afterglows (e.g., GRB 970508 and GRB 000301C).
In our model, these humps are understood to be due to
the contribution of the radiation from regions 2 and 3 when
the reverse shock crosses region 4. A few other interpretations
have been proposed, e.g., refreshed shocks due to Poynting-flux-dominated
or kinetic-energy-dominated injection (Dai \& Lu 1998b; Panaitescu,
M\'esz\'aros \& Rees 1998; Zhang \& M\'esz\'aros 2001a, 2001c;
Chang et al. 2001), and microlensing events (Garnavich, Loeb \&
Stanek 2000). However, there is not any abrupt drop in the afterglow light
curves of the latter models. This property could be used to distinguish
between the present model and the other interpretations. We will carry out
detailed fits to the optical afterglows of GRB 970508 and GRB 000301C
based on the present model. In addition, this model can account well
for the rise-decline feature of the X-ray afterglow light curve of 
GRB 981226 (Frontera et al. 2000).  

Finally, we have noted that when the reverse shock crosses 
region 4, neutrinos with energies of TeV-PeV are possibly produced by
$\pi^+$ created in interactions between accelerated protons
and synchrotron photons from accelerated electrons
in regions 2 and 3. Furthermore, such neutrino emission is delayed
about $t_{\Delta,3}$ after the GRB. This result is different from the prompt
neutrino emission discussed by many authors, e.g, Waxman \& Bahcall (1997,
2000), Bahcall \& M\'esz\'aros (2000), M\'esz\'aros \& Rees (2000),
Dai \& Lu (2001), and M\'esz\'aros \& Waxman (2001). One expects that
these delayed neutrinos, if detected, could provide further diagnostics
about circumburst density-jump environments.

%%%%%%%%%%%%%%%%%%%%%%%%%%%%%%%%%%%%%%%%%%%
 \acknowledgments

We thank Ralph Wijers for  valuable discussions, and Y. F. Huang, 
X. Y. Wang, B. Zhang  and the referee for useful comments. We particularly 
thank B. Zhang for careful reading the manuscript. This work was supported 
by the National Natural Science Foundation of China (grants 19825109 
and 19773007) and the National 973 Project (NKBRSF G19990754).

\clearpage
\begin{figure}
\begin{picture}(100,250)
\put(0,0){\includegraphics{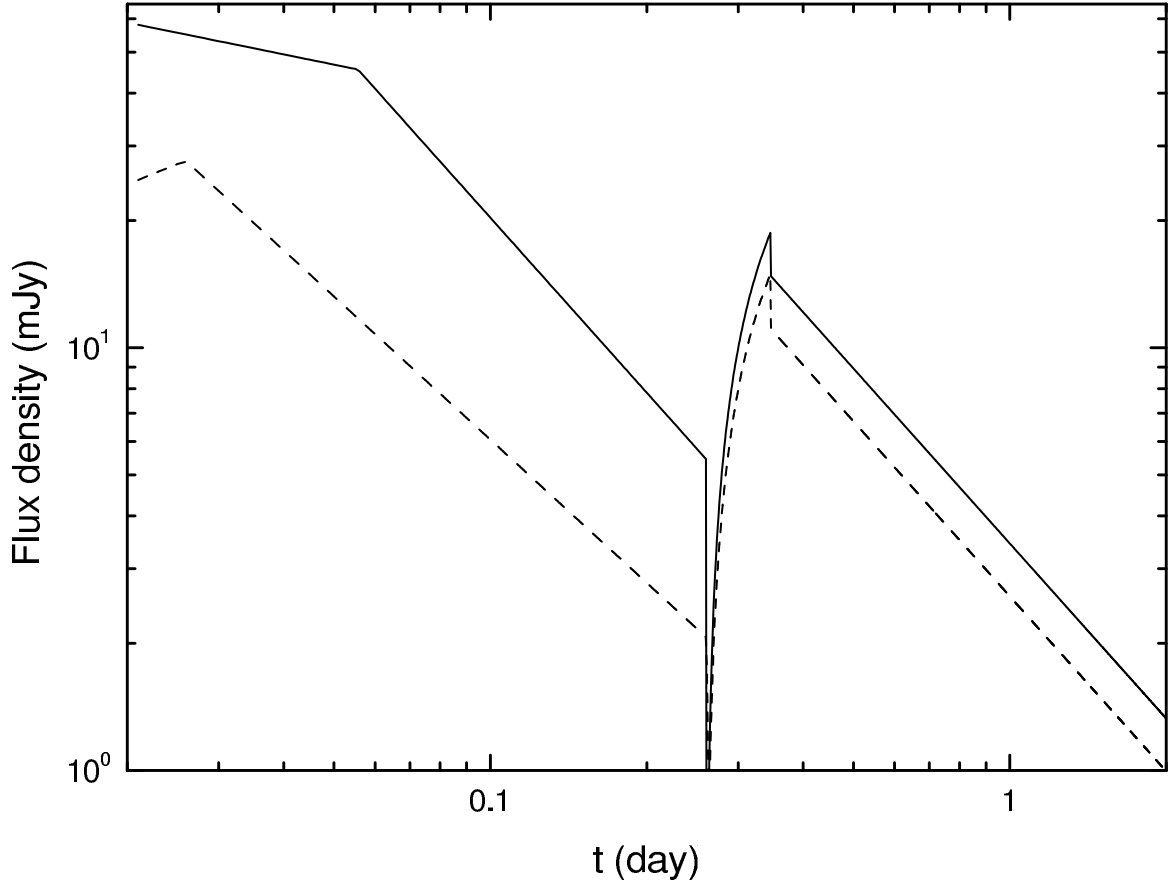}}
\end{picture}
\caption
{R-band light curves of the afterglow emission when an
ultrarelativistic blast wave interacts with a density-jump medium.
The blast wave expands within an ISM ($s=0$) until it reaches a
high-density medium at $R_0=5\times 10^{17}$ cm (or $t_0=0.26$
days). The model parameters are taken: $E_{53}=1$, $n_0=1$,
$n_{1,3}=1$, $\epsilon_e=0.1$, $p=2.5$ and $D_L=2\times 10^{28}$
cm. The solid and dashed lines correspond to $\epsilon_B=0.1$ and
$0.01$ respectively.} 
\end{figure}

\clearpage
\begin{figure}
\begin{picture}(100,250)
\put(0,0){\includegraphics{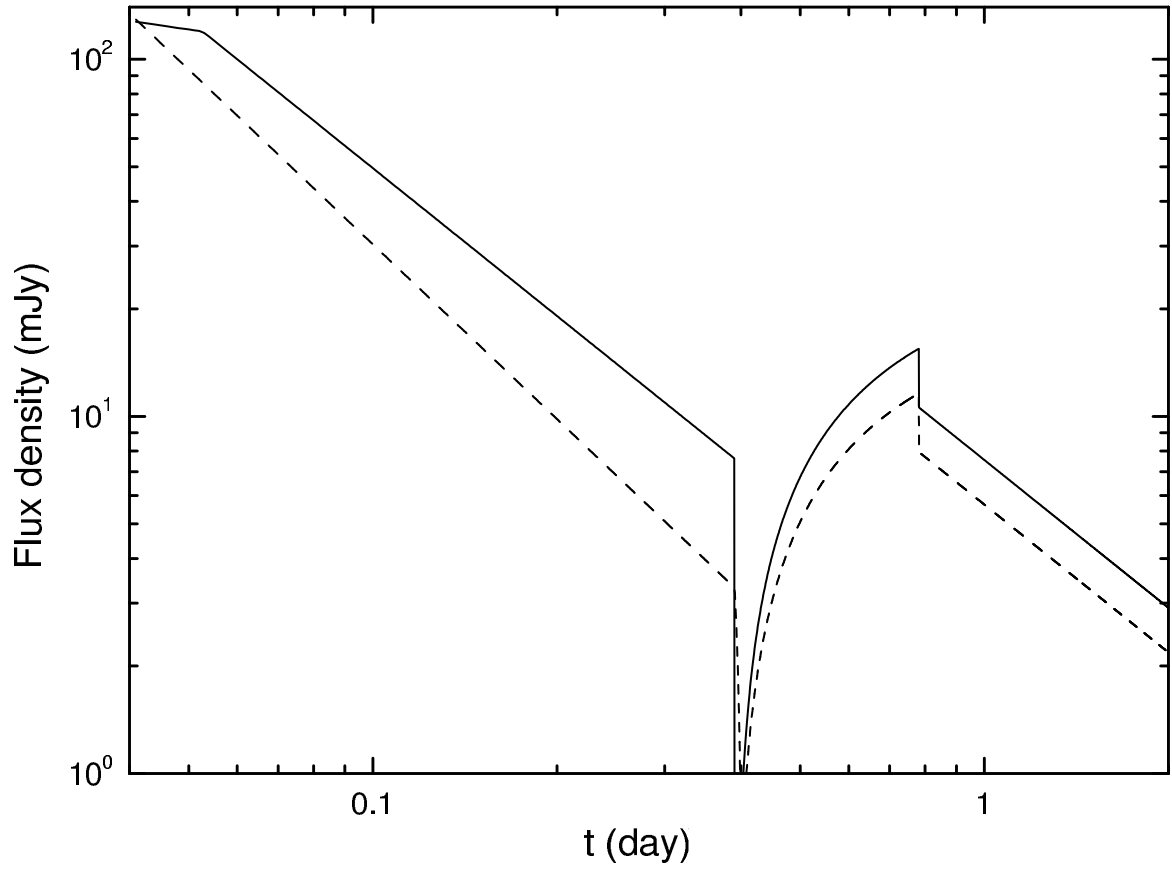}}
\end{picture}
\caption
{Same as Fig. 1 but the inner medium at $R\le R_0=5\times 10^{17}$ cm 
is a stellar wind ($s=2$) with $A_*=1$ and $t_0=0.39$ days.  }
\end{figure}

\end{document}